\documentclass[showpacs,aps,prb,twocolumn,floatfix]{revtex4}
\usepackage{times}
\usepackage{graphicx}
\usepackage{bm}
\usepackage{amsmath}
\usepackage{amssymb}

\newcommand{\s}{\sum\limits}
\newcommand{\p}{\prod\limits}

\newcommand{\be}{\begin{equation}}
\newcommand{\e}{\end{equation}}
\newcommand{\beml}{\begin{subequations}}
\newcommand{\eml}{\end{subequations}}
\newcommand{\beq}{\begin{eqnarray}}
\newcommand{\eq}{\end{eqnarray}}
\newcommand{\ba}{\begin{array}}
\newcommand{\ea}{\end{array}}
\newcommand{\lt}{\left}
\newcommand{\rt}{\right}
\newcommand{\n}{\nonumber}
\newcommand{\la}{\langle}
\newcommand{\ra}{\rangle}

\newcommand{\bb}{\boldsymbol}

\newcommand{\da}{\downarrow}
\newcommand{\ua}{\uparrow}
\newcommand{\h}{^\dagger}

\DeclareMathOperator{\tr}{Tr}

\begin{document}

\title{Finite-temperature Bell test for quasiparticle entanglement in the Fermi sea}
\author{W.-R.~Hannes}
\affiliation{Department of Physics, University of Konstanz, D--78457 Konstanz, Germany}
\author{M.~Titov}
\affiliation{School of Engineering \& Physical Sciences, Heriot-Watt University, Edinburgh EH14 4AS, UK}
\date{February 2008}

\begin{abstract}
We demonstrate that the Bell test cannot be realized 
at finite temperatures in the vast majority 
of electronic setups proposed previously for quantum entanglement generation. 
This fundamental difficulty is shown to originate in a finite probability 
of quasiparticle emission from Fermi-sea detectors. 
In order to overcome the feedback problem we suggest a detection strategy, 
which takes advantage of a resonant coupling to the quasiparticle drains.
Unlike other proposals the designed Bell test provides 
a possibility to determine the critical temperature 
for entanglement production in the solid state. 
\end{abstract}


\maketitle
It is well-known that, unlike photons, 
quasiparticles in the Fermi sea injected from reservoirs, 
which are kept at thermal equilibrium, can be entangled 
by just a tunnel barrier.\cite{Bee03}
This allows for particularly simple proposals for quantum 
quasiparticle entanglement, which do not involve 
interactions.\cite{Bee03,Bee04,Sam04a,Leb05,Lor05} 
Theoretical results for the entanglement production 
in different electronic setups have been summarized 
in Refs.~\onlinecite{Sam04b}~and~\onlinecite{Bee05a}, while yet no experimental 
evidence of the quasiparticle entanglement 
in the Fermi sea has become available. 

The quantum entanglement of two particles with respect 
to a spin-like degree of freedom can be accessed 
experimentally by measuring the spin correlator 
\be
\label{C1}
{\cal C}({\mathbf a},{\mathbf b})=\lt\la
\lt(\mathbf{a}\cdot\bb{\sigma}\rt)_1\otimes
\lt(\mathbf{b}\cdot\bb{\sigma}\rt)_2
\rt\ra,
\e
where $\bb{\sigma}=(\sigma_x,\sigma_y,\sigma_z)$ is the vector
of Pauli matrices. The spin projection of the particles 
in the detectors $1$ and $2$ is measured with respect 
to the unit vectors $\mathbf{a}$ and $\mathbf{b}$, correspondingly.
If the correlation between the particles is of a classical 
origin the following Bell inequality holds,\cite{Cla69}
\be
\label{B}
{\cal B}=|{\cal C}({\mathbf a},{\mathbf b})+{\cal C}({\mathbf a}',{\mathbf b})
+{\cal C}({\mathbf a},{\mathbf b}')-{\cal C}({\mathbf a}',{\mathbf b}')|\leq 2,
\e
for arbitrary choice of the unit vectors
$\mathbf{a}$, $\mathbf{b}$, $\mathbf{a}'$, $\mathbf{b}'$.
The violation of the inequality (\ref{B}) is, therefore,  
sufficient but not necessary condition for quantum entanglement.

In solid-state electronics we deal with 
elementary excitations in the Fermi gas, which are referred 
to as quasiparticles. Even though the pairwise quasiparticle 
entanglement\cite{Bur00,Bur03} is believed to be 
generated in many devices,\cite{Sam03,Sca04,Bee05b,Sam05}
its experimental observation is obscured by the nature of 
electronic detectors. Those, unlike the photodetectors 
in optical setups, contain a number of quasiparticles in 
the ground state, which fill up available quantum levels 
below the Fermi energy. If a part of the device 
is at finite temperature the electron and hole 
excitations are spontaneously created near the Fermi surface
resulting in a finite probability for a Fermi-sea detector 
to emit. Such processes are harmful for any sensible Bell test. 

The problem of quasiparticle entanglement detection
has been put forward in Refs.~\onlinecite{Kaw01}~and~\onlinecite{Cht02}, 
where the possibility to construct a Bell-type inequality 
with current cross-correlators is discussed. 
It has been suggested to take advantage of the generalized 
spin-correlator
\be
\label{C2}
{\cal C}^{\mathrm M}({\mathbf a},{\mathbf b}) =\frac{\lt\la\lt(N_{1\ua}-N_{1\da}\rt)\lt(N_{2\ua}-N_{2\da}\rt)\rt\ra}
{\lt\la\lt(N_{1\ua}+N_{1\da}\rt)\lt(N_{2\ua}+N_{2\da}\rt)\rt\ra},
\e
where $N_{n\sigma}$ is a number of particles with a spin projection 
$\sigma$ registered by the detector $n$. (In solid state the role of spin 
can be played by other quantum degrees of freedom such as orbital momentum or isospin). 
Similarly to Eq.~(\ref{C1}) the spin projection in Eq.~(\ref{C2}) is measured with respect 
to the direction $\mathbf{a}$ in the first detector and $\mathbf{b}$ in the second one.
Both definitions (\ref{C1}) and  (\ref{C2}) are equivalent 
in the original Bell setup, if no more than two particles are 
received within the detection time
and the detectors do not emit particles.
In electronic circuits 
the number of quasiparticles 
$N_{n\sigma}$ is given by the time integral of a current $I_{n\sigma}$ flowing to the 
corresponding Fermi-sea reservoir
\be
\label{N}
N_{n\sigma}\propto \int_0^{t_{\rm det}}\!\!\!\! dt\, I_{n\sigma}(t),
\e
which is not restricted. For large detection times $t_{\rm det}$
one typically observes $|N_{n\ua}-N_{n\da}|\ll |N_{n\ua}+N_{n\da}|$, hence 
the Bell inequality (\ref{B}) cannot be violated and the corresponding measurement 
is useless for entanglement detection. The difficulty has been discussed 
in Ref.~\onlinecite{Cht02} for zero temperature.

An essential problem occurs in the opposite limit  
$t_{\rm det}\to 0$, because $N_{n\sigma}$ defined by Eq.~(\ref{N})
can take on negative values. This leads to fluctuations with 
$|N_{n\ua}-N_{n\da}| > |N_{n\ua}+N_{n\da}|$, which are explicitly forbidden 
in the Bell test. This situation is realized at finite temperatures.
Then, the violation of Eq.~(\ref{B}) has no relation to the entanglement 
detection and the corresponding measurement is not of a Bell type. 

\begin{figure}[tb]
\includegraphics[width=0.9\linewidth]{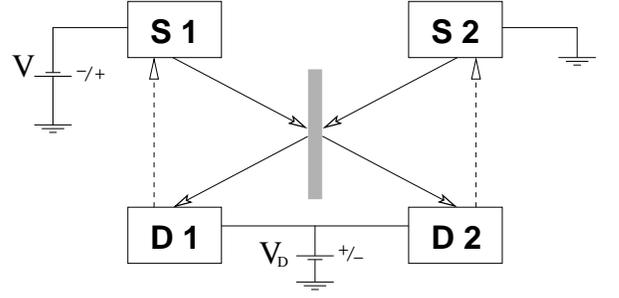}
\caption{A generic beam splitter for entanglement production in the solid state.
The voltage bias applied between the sources $S1$ and $S2$ generates an entangled 
outgoing state at the detectors $D1$, $D2$
provided the temperature in the sources $T$ is smaller than 
a critical temperature $T_c$. 
}
\label{fig:setup}
\end{figure}

\begin{figure}[tb]
\includegraphics[width=0.75\linewidth]{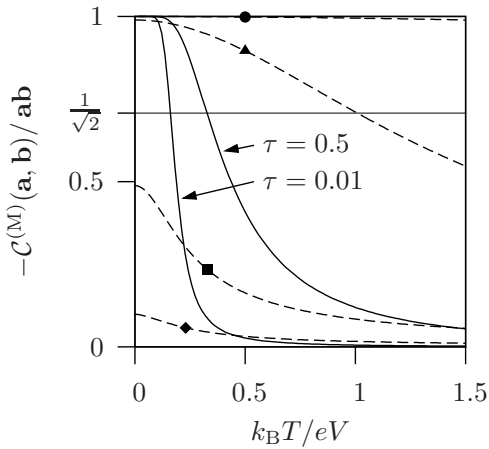}
\caption{The spin cross-correlator $\cal C$ obtained from the density matrix 
of the final scattering state (solid lines; cf.~Eq.~\ref{Cxi}), and
its generalization ${\cal C}^{\mathrm M}$, evaluated numerically from 
Eqs.~(\ref{Cgamma}-\ref{s}) for different values of the detection time 
$e V t_{\rm det}/h =0.01 (\bullet), 0.1 (\blacktriangle), 
1({\scriptstyle\blacksquare}), 5({\scriptstyle\blacklozenge})$ 
(dashed lines; see Eqs.~\ref{useless},\ref{nobell}).
}
\label{fig:result}
\end{figure}

Thus, the violation of the inequality (\ref{B}) with 
the correlator ${\cal C}$ substituted by ${\cal C}^{\mathrm M}$ 
does not provide a conclusive evidence for quantum entanglement 
generation at any finite temperature. This difficulty clearly applies to the 
detection of electron-hole entanglement\cite{Bee05a,Bee05b} produced 
by tunneling events or by time-dependent gating. 
But even in more sophisticated setups where zero-temperature detectors and 
finite-temperature sources are represented by different metallic leads 
(in close resemblance to the original Bell proposal) the Bell test 
based on Eq.~(\ref{C2}) is flawed.
Examples include three-terminal fork geometries\cite{Leb05} and 
four-terminal beam-splitter geometries with grounded detectors. 
We focus on the latter (see Fig.~\ref{fig:setup}) due to a number of 
previously proposed realizations,\cite{Bee03,Sam04a,Leb05,Lor05,Sam03,Gio06,Fao07}
which are mostly based on the directed transport along 
quantum-Hall edge channels. 
Minor modifications, 
such as lowering chemical potential in one of the detectors
or increasing detection time, can suppress the probability 
of detector emission but lead, instead, 
to useless measurement with 
$|N_{n\ua}-N_{n\da}|\ll |N_{n\ua}+N_{n\da}|$. 
The generic situation is illustrated in Fig.~\ref{fig:result}
for the case of electronic beam splitter.

\begin{figure}[tb]
\includegraphics[width=0.75\linewidth]{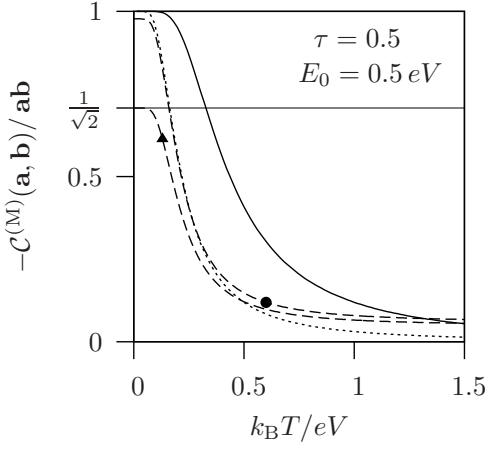}
\caption{
The case of resonant detector coupling. 
The short-dashed line shows ${\cal C}^{\mathrm M}$ from Eq.~(\ref{final}),
while the long-dashed lines are numerical results for Breit-Wigner resonances (\ref{tE}) 
with finite width $\Gamma=0.01eV$, detector voltage $V_D=-V$, 
and different detection times $\Gamma t_{\rm det}/h =0.01(\bullet), 0.1(\blacktriangle)$.
The measurement is useless for $t_{\rm det} \gtrsim 0.1 h/\Gamma$.
The solid line shows the correlator ${\cal C}$ from Eq.~(\ref{Cxi}).
}
\label{fig:proposal}
\end{figure}

For quantum particles the spin correlator from Eq.~(\ref{C2}) 
is expressed through the expectation value
\beq
\label{K}
&&\la N_{1\sigma} N_{2\sigma'}\ra \propto K_{\sigma\sigma'},\\
&&K_{\sigma\sigma'}=t_{\rm det}^2 \lt[\la I_{1\sigma}\ra \la I_{2\sigma'}\ra
+ \int \frac{d\omega}{2\pi}\, {\cal P}_{\sigma\sigma'}(\omega){\cal F}(\omega t_{\rm det}/2)\rt].
\n
\eq
where $N$ and $I$ are regarded as operators. We introduce the function 
${\cal F}(x)=(\sin x)^2/x^2$ and the frequency-dependent cross-correlator
\be
\label{P}
{\cal P}_{\sigma\sigma'}(\omega)=\int dt\, e^{i\omega t}\la \delta I_{1\sigma}(t) I_{2\sigma'}(0) \ra,
\e
with $\delta I_{n\sigma}(t)= I_{n\sigma}(t)-\la I_{n\sigma}\ra$.
In Figs.~\ref{fig:result},\ref{fig:proposal} the correlator ${\cal C}^{\mathrm M}$ defined by 
experimentally measurable quantities (\ref{K},\ref{P}) is 
compared with the exact result of the density matrix analysis 
of the final state.\cite{Bee05a} 

We restrict ourselves to an important class of systems 
which do not involve spin-dependent scattering, because the chances 
to generate quantum entanglement with respect to the spin degree of freedom 
are obviously maximized in such setups. The values of ${\cal P}_{\sigma\sigma'}$ 
in Eq.~(\ref{P}) are related to the cross-correlator 
${\cal P}$ of the corresponding spin-independent problem as 
\beq
\label{P1}
&&{\cal P}_{\ua\ua}={\cal P}_{\da\da}=\tfrac{1}{2}(1+\mathbf{a b}){\cal P}, \\
\label{P2}
&&{\cal P}_{\ua\da}={\cal P}_{\da\ua}=\tfrac{1}{2}(1-\mathbf{a b}){\cal P}.
\eq
This symmetry holds even for interacting electronic systems provided 
the absence of spin dephasing. It follows from Eqs.~(\ref{P1},\ref{P2})
that a neglection of the mean currents\cite{Lor05,Bee05a} 
in Eq.~(\ref{K}) is equivalent to ${\cal C}^{\mathrm M}({\mathbf a},{\mathbf b})=\mathbf{ab}$, hence 
the inequality (\ref{B}) is violated with ${\cal B}_{\rm max}=2\sqrt{2}$
irrespective of voltages, temperature or other setup characteristics. 
Clearly, such violation has nothing to do with pairwise quantum entanglement. 
We will see that the problem persists even if the exact expression 
for $K_{\sigma\sigma'}$ is used.   

In the absence of spin-dependent scattering the mean currents measured 
by the detectors do not depend on the directions $\mathbf{a}$ and $\mathbf{b}$, 
$\la I_{n\sigma} \ra= \la I_n \ra$. From Eqs.~(\ref{K},\ref{P},\ref{P1},\ref{P2}) 
we obtain
\be
\label{Cgamma}
{\cal C}^{\mathrm M}({\mathbf a},{\mathbf b})=\frac{\gamma}{2+\gamma}\,\mathbf{ab},
\e
with a parameter $\gamma$ given by the ratio 
\be
\label{gamma}
\gamma=\frac{\int d\omega\;{\cal P}(\omega) {\cal F}(\omega t_{\rm det}/2)}
{2\pi \la I_1 \ra \la I_2 \ra},
\e
where both the cross-correlator ${\cal P}(\omega)$ and the product of the mean currents 
have to be calculated for the corresponding spin-independent problem.

An example of such a calculation can be performed within 
the Landauer-B\"uttiker scattering approach,\cite{Bla00} 
which is valid as far as 
inelastic processes in between the reservoirs can be disregarded. 
Within the scattering approach the mean current to the reservoir $\alpha$ is given by
\be
\label{scI}
\la I_{\alpha}\ra=\frac{e}{h}\int dE\s_\beta \lt(\delta_{\alpha\beta}-|S_{\alpha\beta}(E)|^2\rt)f_\beta(E),
\e
where $f_{\alpha}(E)=
(1+\exp\lt[(E-eV_\alpha)/k_{\rm B}T_\alpha\rt])^{-1}$ 
is the Fermi distribution function, which
depends on the temperature of the corresponding reservoir $T_{\alpha}$ and
the voltage bias $V_\alpha$ applied. The frequency-dependent correlator 
(\ref{P}) of the currents flowing to the reservoirs $\alpha$ and $\alpha'$ 
reads,\cite{Bla00}
\beq
&&{\cal P}_{\alpha\alpha'}(\omega)=\frac{e^2}{2h}\int dE\;
\s_{\beta\beta'}M_{\alpha\alpha',\beta\beta'}(E,\hbar\omega)F_{\beta\beta'}(E,\hbar\omega),\n\\
&&F_{\beta\beta'}(E,\Omega)=f_\beta(E)\tilde{f}_{\beta'}(E+\Omega)+\tilde{f}_\beta(E)f_{\beta'}(E+\Omega),\n\\
&&M_{\alpha\alpha';\beta\beta'}(E,\Omega) =
\lt(\delta_{\alpha\beta}\delta_{\alpha\beta'}-S^*_{\alpha\beta}(E)S_{\alpha\beta'}(E+\Omega)\rt)\n\\
&&\qquad\quad\times\lt(\delta_{\alpha'\beta}\delta_{\alpha'\beta'}
-S^*_{\alpha'\beta'}(E+\Omega)S_{\alpha'\beta}(E)\rt),
\label{scP}
\eq
\be
\tilde{f}(E)\equiv 1-f(E).
\e

Let us consider a generic beam splitter with no 
spin-dependent scattering depicted schematically in Fig.~\ref{fig:setup}. 
Such a setup is characterized by an energy-independent $S$-matrix
\be
\label{S}
S=\lt(\ba{cc}
0 & s'\\ s & 0
\ea\rt),
\e
where $2\times 2$ unitary matrices $s$ and $s'$ describe the transport 
from sources to detectors and from detectors to sources, correspondingly. We parameterize
\be\label{s}
s=\lt(\ba{cc}
\! e^{i\phi}\!& 0\\ 0& \! e^{i\phi'}\!
\ea\rt)
\lt(\ba{cc}
\!\sqrt{1-\tau}\! & \! i\sqrt{\tau}\! \\ \! i\sqrt{\tau}\! & \! \sqrt{1-\tau}\!
\ea\rt)
\lt(\ba{cc}
\! e^{i\theta}\! & 0\\ 0 & 
\! e^{i\theta'}\!
\ea\rt),
\e
where $\tau\in [0,1]$ is the beam-splitter transparency 
and the spin index is omitted. Following the majority of proposals both 
detectors and the second source are grounded, i.e. $V_D\equiv V_{D1}=V_{D2}=0$, 
$V_{S2}=0$, while $V_{S1}=V$ is the voltage applied between the sources. 

At zero temperature the beam splitter acts\cite{Bee05a} as a source of 
spin-entangled Bell pairs 
\be
\label{bell}
|\Psi_B\ra=\frac{1}{\sqrt{2}}| \ua_1\da_2-\da_1\ua_2\ra
\e 
where the index $n=1,2$ refers to the detector number.
Such an entanglement generation is due to the Pauli principle, 
which guarantees that a filled state with $E\in (0,eV)$ 
in the first source 
contains exactly two quasiparticles with the opposite spins.

The Bell pairs can be accessed at zero temperature
by performing a time coincidence detection.
For finite temperature $T$ in the sources 
the density matrix projection, which corresponds 
to a single particle in each detector, 
is derived in the Appendix~\ref{app:dm}\cite{Bee05a}
\be
\label{rho}
\rho^{\rm out}_{11}=\tfrac{1}{4}(1-\xi)\openone_4 + \xi|\Psi_B \ra\la \Psi_B|,
\e
where $\openone_4$ is the unit matrix in the two-particle Hilbert space 
and $\xi$ is an energy-independent weight factor
\beq
\label{xi}
\xi=\frac{\tau(1-\tau)(f_{S1}-f_{S2})^2}{\tau(1-\tau)(f_{S1}-f_{S2})^2
+2f_{S1}\tilde{f}_{S1}f_{S2}\tilde{f}_{S2}},
\eq
where $f_{Sn}$ is the Fermi distribution function in $n$-th source.
The result (\ref{rho}) describes the mixed Werner state,\cite{Wer89} 
which is entangled as far as $\xi>1/3$ according 
to the Wootters formula.\cite{Woo98} 
In the present case this 
condition is equivalent to $T<T_c$ with the critical temperature $T_c$ 
determined by the equation\cite{Bee05a}
\be
\label{Tc}
\tau(1-\tau)\sinh^2(eV/2k_{\rm B}T_c)=1/4.
\e
From Eqs.~(\ref{C1},\ref{rho}) one obtains 
the exact spin correlator
\be
\label{Cxi}
{\cal C}({\mathbf a},{\mathbf b})=-\xi\, \mathbf{ab},
\e
which is plotted in Fig.~\ref{fig:result} with the solid line
for different values of the transparency parameter.
The corresponding Bell inequality (\ref{B}) can be violated 
for $\xi>1/\sqrt{2}$, which is, indeed, a sufficient condition for the entanglement.
Whether or not such a Bell test can be performed by measuring 
current cross-correlator (\ref{C2}) is, however, an open question.

In order to answer this question
we substitute the expression (\ref{S}) for the $S$-matrix 
to Eqs.~(\ref{scI},\ref{scP}), where 
the summation runs over the index $\alpha=\{S1,S2,D1,D2\}$.
The correlator ${\cal C}^{\mathrm M}$ is, then, obtained
from Eqs.~(\ref{Cgamma},\ref{gamma}) with $I_1\equiv I_{D1}$, $I_2\equiv I_{D2}$, 
and ${\cal P}\equiv {\cal P}_{D1,D2}$.

For $t_{\rm det}\gg {\rm min}\{h/eV,h/k_{\rm B}T\}$ we obtain
\be
\label{useless}
\gamma = -\frac{h}{eVt_{\rm det}}\lt(\coth\lt(\frac{eV}{2k_{\rm B}T}\rt)- \frac{2k_{\rm B} T}{eV}\rt)\ll 1, 
\e
i.e. the corresponding measurement is useless for an entanglement detection.
Indeed, such a long-time measurement is not projective, therefore
it does not single out the state with one quasiparticle in each detector.\cite{Bay06} 

In the opposite limit we, however, find 
\be
\label{nobell}
\gamma = - 1, \qquad t_{\rm det}\ll {\rm min}\{h/eV,h/k_{\rm B}T\},
\e
hence the inequality (\ref{B}) is violated for any temperature 
of the source. Thus, according to the density matrix analysis 
(\ref{rho},\ref{Cxi}), the corresponding measurement is not of a Bell type.
Both results (\ref{useless}) and (\ref{nobell}) formally hold
for any temperature of the detectors.
 
The transition from non Bell-type measurement 
to the useless measurement with the increase of $t_{\rm det}$
is illustrated in Fig.~{\ref{fig:result}}. 
The Bell parameter defined with the correlator ${\cal C}^{\mathrm M}$
does not depend on the beam-splitter transparency $\tau$
and can easily exceed $2$ even in the absence of any entanglement.  

The result of Eq.~(\ref{nobell}) is equivalent to 
\be
\label{explain}
\la I_{D1}(t) I_{D2}(t) \ra=0.
\e
At $T=0$ the currents $I_{Dn}(t)$ are sign-definite, hence 
Eq.~(\ref{explain}) is exact for every single time-coincidence measurement
in agreement with the prediction of the density matrix approach.
For rising temperatures $T>0$ the correlation (\ref{explain})
holds only on average and is not sensitive to vanishing 
quantum entanglement in the final state of the beam splitter (\ref{rho}).
Consequently, the inequality (\ref{B}) with ${\cal C}$ substituted by ${\cal C}^{\mathrm M}$
can be violated for arbitrarily high temperatures. The absence of critical temperature 
indicates once again\cite{Fin05} that such a violation has nothing
to do with the entanglement detection. Instead, the decay of 
${\cal C}^{\mathrm M}({\mathbf a},{\mathbf b})$ with the temperature 
in Fig.~\ref{fig:result} (dashed lines) is determined by the detection time $t_{\rm det}$.

Thus, the measurement of ${\cal C}^{\mathrm M}({\mathbf a},{\mathbf b})$ 
cannot be used for the entanglement test in the beam-splitter setup and  
the value of $T_c$ cannot be inferred from such a measurement as 
the matter of principle. 

We propose a way to rescue the Bell measurement by 
coupling detectors via the energy filters, which 
are described by energy dependent scattering amplitudes:
$r_n,r'_n,t_n,t'_n$, where $n=1,2$ is the number of the detector. 
The use of energy-filters in the context of Bell measurement at 
zero temperature has been discussed in Ref.~\onlinecite{Les01}.
Let us illustrate our results for the case of identical filters 
with the Breit-Wigner form of the transmission amplitude
\be
\label{tE}
t_n(E)=e^{i\delta_n}(\Gamma/2)(E-E_0-i\Gamma/2)^{-1}.
\e
The $S$-matrix of the full setup including 
the filters is given by
\be
\label{SE}
S(E)=\lt(\ba{cc} 
s'r(E)\,s& s't'(E)\\ t(E)\,s & r'(E)
\ea\rt)
\e
where $t={\rm diag}(t_1,t_2)$,  $r={\rm diag}(r_1,r_2)$, etc. 
The condition for time-coincidence detection now reads
$t_{\rm det}\ll h/\Gamma$.  The currents $I_{Dn}(t)$ can 
be made sign-definite by applying an additional voltage bias 
$V_D$, as shown in Fig.~\ref{fig:setup}. The current fluctuations 
due to temperature are not harmful for the Bell test 
as far as $|eV_D|\gg k_{\rm B}T$, which is the only restriction on 
the value of $V_D$. In this case there is no 
requirement for an additional cooling of the detectors, so that a whole
setup can be kept in temperature equilibrium.
Moreover, for $\Gamma \ll eV$ the dependence on $t_{\rm det}$ vanishes, 
meaning that ${\cal C}^{\mathrm M}({\mathbf a},{\mathbf b})$ can be obtained 
experimentally from zero-frequency noise measurements.
The feasibility of such a Bell test is illustrated in Fig.~\ref{fig:proposal}
for realistic values of the parameters.

For $\Gamma\to 0$ we obtain from Eqs.~(\ref{Cgamma},\ref{gamma},\ref{scP},\ref{tE},\ref{SE})
\be
\label{final}
{\cal C}^{\mathrm M} =-\lt.\frac{\tau(1-\tau)(f_{S1}-f_{S2})^2\,\mathbf{ab}}{\tau(1-\tau)(f_{S1}-f_{S2})^2
+2f_{S1}f_{S2}}\rt|_{E=E_0}.
\e
The result is plotted with the short-dashed line in Fig.~\ref{fig:proposal}. 
It is evident from the comparison with Eqs.~(\ref{xi},\ref{Cxi}) 
that the proposed measurement is always of the Bell type. 
The correlator (\ref{final}) tends to the exact one (\ref{Cxi})
for $E_0 \gg eV$. The setup efficiency is, however, 
exponentially low in this limit. The numerical results in the case of finite resonance 
width $\Gamma=0.01 eV$ are plotted in Fig.~{\ref{fig:proposal}} 
with the dashed lines. The test provides the lower estimate for the critical temperature.

In conclusion we point out the fundamental restrictions for 
the Bell test in electronic setups due to the quasiparticle 
emission from Fermi-sea detectors. We propose a way to rescue 
the Bell measurement by a resonant coupling to the detectors.
We show that the lower estimate of the critical temperature 
for entanglement production can be experimentally obtained 
in the proposed setup.

This research was supported by the DFG Priority Programm 1285.
The discussions with C.~W.~J.~Beenakker, W.~Belzig, and Yu.~V.~Nazarov
are gratefully acknowledged.

\appendix

\section{Density matrix projection} \label{app:dm}

Following Ref.~\onlinecite{Bee05a}
we review the derivation of the density matrix
projection (\ref{rho}) for final scattering state 
in the case of the setup depicted in Fig.~\ref{fig:setup}.
The density matrix of the incoming state is given by
\be
\label{rho-in}
\rho^{\rm in} = \p_{n,E,\sigma} 
\lt( \tilde{f}_{Sn}(E) \,|0 \ra\la 0|
+f_{Sn}(E)\, a_{n\sigma E}\h \,|0\ra\la 0|\, a_{n\sigma E}
\rt),
\e
where $f_{Sn}$ is the Fermi distribution function in the 
source $S_n$, $\tilde{f}_{Sn}\equiv 1-f_{Sn}$, 
and $a_{n\sigma E}$ 
is the fermion annihilation operator for
an incoming scattering state at the channel $n$ and energy $E$. 
The annihilation operators for the outgoing 
scattering states, $b_{n\sigma E}$, are obtained 
from the relation,
\be\label{b}
b_{n\sigma E}=\s_m s_{nm}(E)\, a_{m\sigma E},
\e
where $s_{nm}$ are the components of a unitary scattering matrix.

Thus, the density matrix of the final state is
\be
\label{rho-out}
\rho^{\rm out}= \p_{n,E,\sigma} \lt\{\tilde f_{Sn}(E)\,|0\ra\la0|+
f_{Sn}(E)\,c_{n\sigma E}\h\,|0\ra\la0| c_{n\sigma E}\rt\},
\e
where
\be
c_{n\sigma E}\h=\sum_m  b_{m\sigma E}\h\, s_{mn}(E).
\e

In order to quantify the two-particle entanglement for the partition 
${\cal H}_{D1}\otimes{\cal H}_{D2}$ of Hilbert space with respect 
to the detectors, the state $\rho^{\rm out}$ has to be projected 
onto the sectors ${\cal N}_{E_1 N_1,E_2 N_2}$ of the Fock space 
with the energies $E_1, E_2$ and
particle numbers $N_1, N_2$ in the 
corresponding detectors $D_1, D_2$. 
The density matrix $\rho_{N_1,N_2}^{\rm out}$ of the projection 
factorizes into a product state in all sectors except 
for the sector ${\cal N}_{E 1,E 1}$ with $E_1=E_2=E$ and $N_1=N_2=1$.
Projection onto this sector is found from Eq.~(\ref{rho-out}) as  
\beq
\n
w_{11}\rho_{11}&=&f_{S1}\tilde f_{S1}f_{S2}\tilde f_{S2}\ \openone_4\\
&+&2\tau(1-\tau)(f_{S1}-f_{S2})^2|\Psi_B\ra\la\Psi_B|, 
\label{rho11}
\eq
where $\openone_4$ is the unit matrix in the two-particle Hilbert space,
$|\Psi_B\ra$ is the Bell state (\ref{bell}), and the weight 
factor $w_{11}$ is determined from the condition $\tr\rho_{11}=1$ as
\be
\label{w11}
\quad w_{11}= 4 f_{S1}\tilde f_{S1}f_{S2}\tilde f_{S2} +
2\tau(1-\tau)(f_{S1}-f_{S2})^2.
\e
From Eqs.~(\ref{rho11},\ref{w11}) we obtain Eqs.~(\ref{rho},\ref{xi}).
By substituting $f_{Sn}=(1+\exp[(E-eV_n)/k_BT])$ with $V_1=V$, $V_2=0$
in Eq.~(\ref{xi}) we can further simplify the parameter $\xi$ as
\be\label{xi1}
\xi(T)=1-\lt[1+2\tau(1-\tau)\,\sinh^2\frac{eV}{2k_BT}\rt]^{-1}.
\e
The critical temperature $T_c$ is determined from the equation 
$\xi(T_c)=1/3$, which is equivalent to Eq.~(\ref{Tc}).

\section{Evaluation of the correlator ${\cal C}^M$}
\subsection{Plain beam splitter}\label{app:bs}

We evaluate the generalized spin-correlator 
${\cal C}^{\mathrm M}({\mathbf a},{\mathbf b})$ 
given by Eqs.~(\ref{Cgamma},\ref{gamma}) 
in the framework of the scattering approach.
By substituting the scattering matrix (\ref{S}, \ref{s}),
into Eq.~(\ref{scI}) we calculate the mean currents,
which are measured in the detectors $D_1, D_2$, as  
\be\label{bsI}
\la I_{D1} \ra =-\,\frac{e}{h}\,(1-\tau)\,eV,\quad
\la I_{D2} \ra =-\,\frac{e}{h}\,\tau\, eV.
\e
The cross-correlator (\ref{scP}) is found as
\beq
{\cal P}_{D1,D2}(\omega)
&=&\frac{e^2}{2h}\,\tau(1-\tau)\lt[2\hbar\omega\coth\lt(\frac{\hbar\omega}{2k_{\rm B} T}\rt)\rt. \\
&-& \s_{\zeta=\pm 1}
\lt.(eV+\zeta\hbar\omega)\coth\lt(\frac{eV+\zeta\hbar\omega}{2k_{\rm B} T}\rt)\rt].\n
\label{bsP}
\eq
The parameter $\gamma$ given by Eq.~(\ref{gamma})
can be calculated analytically in two opposite limits:

(i) For large detection times, $t_{\rm det}\gg {\rm min}\{h/eV,h/k_\mathrm B T\}$,
one can replace $t_{\rm det}{\cal F}(\omega t_{\rm det}/2)$ with 
$2\pi\delta(\omega)$, hence
\be
\gamma=\frac{{\cal P}_{D1,D2}(0)}
{t_{\rm det} \la I_{D1} \ra \la I_{D2} \ra}.
\e
This leads to the result (\ref{useless}).

(ii) For short detection times $t_{\rm det}\ll {\rm min}\{h/eV,h/k_\mathrm BT\}$
one can approximate ${\cal F}(\omega t_{\rm det}/2)\approx 1$ 
in the relevant frequency range $|\hbar\omega|\lesssim eV$.
In this limit the integral in Eq.~(\ref{gamma}) 
does not depend on temperature 
\be
\int d\omega\,{\cal P}_{D1,D2}(\omega) = 
-2\pi \lt(\frac{e^2V}{h}\rt)^2 \tau(1-\tau),
\e
which leads to the simple result (\ref{nobell}). 

\subsection{Beam splitter with energy filters}\label{app:ef}

We repeat the calculation in a more general case of 
an energy-dependent scattering matrix (\ref{SE}). 
From Eq.~(\ref{scI}) we obtain the mean currents 
\beq
\la I_{D1}\ra
&=& \frac{e}{h}\int dE\ |t_1(E)|^2\n\\
&\times& \lt[f_D(E)-(1-\tau)f_{S1}(E)-\tau f_{S2}(E)\rt] ,\label{efI1}\\
\la I_{D2}\ra
&=& \frac{e}{h}\int dE\ |t_2(E)|^2\n\\
&\times& \lt[f_D(E)-\tau f_{S1}(E)-(1-\tau) f_{S2}(E)\rt] ,\label{efI2}
\eq
where $f_{D}(E)=\lt(1+\exp{[(E-eV_D)/k_{\mathrm B}T]}\rt)$ is the Fermi distribution 
function in the detectors. From Eq.~(\ref{scP}) we calculate 
the cross-correlator 
\beq
&&{\cal P}_{D1,D2}(\omega)=\frac{e^2}{2h}\tau(1 -\tau) \int dE \n\\
&&\qquad \times\, t_1^*(E)\,t_1(E+\hbar\omega) \,t_2^*(E+\hbar\omega)\,t_2(E) \n\\
&& \qquad \times\, \big[F_{S1,S1}(E,\hbar\omega)+F_{S2,S2}(E,\hbar\omega) \n\\
&& \qquad\quad -F_{S1,S2}(E,\hbar\omega)-F_{S2,S1}(E,\hbar\omega)\big],
\label{efP}
\eq
where the function $F_{\alpha\beta}$ is defined in Eq.~(\ref{scP}).
These expressions allow for the numerical evaluation of 
${\cal C}^{\mathrm M}({\mathbf a},{\mathbf b})$ for 
arbitrary energy-dependent scattering matrix (\ref{SE}). 

In the case of sharp resonances, such as those of the 
Breit-Wigner form (\ref{tE}) with $\Gamma\to 0$,
we have
\be
\lim_{\Gamma\to 0} \Gamma^{-1}|t_n(E)|^2
=\frac{\pi}{2}\delta(E-E_0),
\label{tEdelta}
\e
and obtain from Eqs.~(\ref{efI1},\ref{efI2},\ref{efP},\ref{tEdelta}) 
\be
\label{efgamma}
\gamma=\frac{-\tau(1-\tau)\lt(f_{S1}-f_{S2}\rt)^2}
{[f_D\!-(1-\tau)f_{S1}\!-\tau f_{S2}]
[f_D\!-\tau f_{S1}\!-(1-\tau)f_{S2}]},
\e
where the Fermi functions are evaluated at the position
of the resonance $E=E_0>0$. For $|eV_D|\gg k_{\mathrm B}T$ 
the value $f_D(E_0)$ is exponentially small, hence 
Eq.~(\ref{final}) is justified. In general the setup 
is functional provided the detector voltage $V_D$ is
sufficiently large to ensure that $f_D$ 
is much smaller than both $f_{S1}$ and $f_{S2}$
within the energy window of the filters.

\end{document}